# Magnetic studies of Ca$_{1-x}$M$_x$RuO$_3$ (M=La and Sr)

I. Felner and U. Asaf

*The Racah Institute of Physics, The Hebrew University of Jerusalem  91904, Israel*
.

CaRuO$_3$ is a perovskite with an orthorhombic distortion and shows the characteristics of spin-glass behavior below T$_C$=87 K. The La$^{3+}$ substitution for Ca$^{2+}$ in Ca$_{1-x}$La$_x$RuO$_3$ samples, induces a disorder in the Ca site (the A site) and the system becomes antiferromagnetically (AFM) ordered with T$_N$ = 58 and 19 K for x=0.1 x=0.5 respectively.  In the Ca$_{1-x}$Sr$_x$RuO$_3$ system, the Ca$_{0.8}$Sr$_{0.2}$RuO$_3$ sample is canted-AFM ordered at 107 K. The compounds with higher Sr concentration display ferromagnetic behavior and the saturation moment increases with Sr. Huge magnetic hysteresis loops are obtained at low temperatures. The coercive field (H$_C$) decreases with Sr. For  x=0.4  H$_C$ =9.5 kOe (at 5 K) whereas  for SrRuO$_3$ H$_C$ =2.4 kOe. For x=0.4  and  0.6,  H$_C$ decreases sharply with temperature and than increases again with a peak at 95 and 115 K, respectively. On the other hand, for SrRuO$_3$, H$_C$ remains practically unchanged up to 50 K and shows a peak at 90 K and than decrease sharply up to T$_C$ =165 K.

## 1. Introduction

Ternary ruthenates exhibit a wide range of electronic and magnetic properties, ranging from superconductivity to ferromagnetism. One class of oxides that has attracted renewed interest is the family of orthorhombic perovskite $CaMO_3$ (M=Ru and Fe) compounds, due to their unusual magnetic properties. The magnetic ground state of $CaRuO_3$ ($Ru^{4+}$, $4d^4:t_{2g}^4 e_g^0$ with S=1) is controversial. Recent publications indicate paramagnetic behavior (or exchange enhanced paramagnetism) down to 30 mK, which is also supported by the single line shape of a $^{99}Ru$ Mössbauer spectrum measured at 4.1 K [1]. The high and low temperature resistivity results indicate that $CaRuO_3$ is a non-Fermi liquid metal [2]. On the other hand, based on magnetic studies of ceramic and single crystal $CaRuO_3$ samples, we have demonstrated that irreversibility appears in the zero-field-cooled (ZFC) field-cooled (FC) magnetization curves, only when measured at low applied magnetic fields [3]. For ceramic samples, a very small hysteresis loop opens at 5 K, and the remanent magnetization of 1.6 emu/mol decreases with temperature and disappears at $T_C \sim$ 87 K. $T_C$ is determined as the temperature at which the remanent magnetization disappears. Mössbauer effect studies (MS) of 0.5% $^{57}Fe$ doped in $CaRuO_3$ show a magnetic sextet at 4.1 K, which disappears at $T_C$. We have proposed that $CaRuO_3$ is not paramagnetic, but rather shows the characteristics of short-range magnetic interactions, possibly spin-glass behavior, or alternatively, the data are consistent with an itinerant electron picture of ferromagnetism. Neutron diffraction experiments at 10 K showed that no additional magnetic neutron scattering peaks exist, indicating that $CaRuO_3$ is not a canted antiferromagnet [4].

It is well established that the isostructural $SrRuO_3$ is a ferromagnet (FM) with a Curie temperature $T_C$ = 165 K. The stark contrast between $SrRuO_3$ and $CaRuO_3$ is surprising because the closed shell s-like character of Sr and Ca do not contribute to the density of states at the Fermi surface and therefore should not be the origin for the different magnetic ground states of these two compounds. Since a common structural feature of the two compounds is that they are composed of an array of corner-shared $RuO_6$ octahedra, it is assumed that the degree of tilting and rotation of these octahedra within their ideal cubic-perovskite structure governs the observed differences in the magnetic ground states. A narrow itinerant 4d band is formed through hybridization of Ru $t_{2g}$ and O 2p orbitals. The 4d bandwidth thus formed sensitively depends on degree of hybridization [5]. $CaRuO_3$ is believed to have a narrow itinerant 4 d-band width, (narrower than for $SrRuO_3$), which is too narrow for long range magnetic ordering, but not so narrow as to cause $CaRuO_3$ to be non-metallic [6]. It means that $CaRuO_3$ is on the verge of magnetic ordering and readily evolves into an ordered phase. On the other hand, no long-range magnetism has been observed in the isostructural orthorhombic $LaRuO_3$ down to 4.2 K [7-8]. Here, $Ru^{3+}$ ($4d^5$, $t_{2g}^5 e_g^0$) ions are assumed to be in the low-spin state with (S=0.5). In that sense $LaRuO_3$ differs from both $CaRuO_3$ and $SrRuO_3$ compounds.

A powerful tool to study physical properties of such systems is through chemical substitution on each Ca and/or Ru sites. Indeed, 5 at % of $Sr^{2+}$, or $Na^+$ substitution for Ca induces anti-ferromagnetic (AFM) or spin glass ordering at T=10 and 55 K respectively [9-10]. Magnetic studies of polycrystalline $CaRu_{1-x}Ti_xO_3$ (0<x<0.1) reveal a ferromagnetic transition at $T_C$=34 K, regardless of the nonmagnetic Ti concentration [11-12]. Substitution for Ru in $CaRu_{1-x}M_xO_3$ (M= Sn and Pb) induces a complicated magnetic structure (probably spin-glass behavior) [13]. Substitution of nonmagnetic $Ti^{4+}$ and $Sn^{4+}$ for $Ru^{4+}$ represents only lattice frustration and magnetic dilution of the Ru sublattice, whereas $Rh^{4+}$ (S=1/2) or $Rh^{3+}$(S=0) ions act as magnetic impurities (different S), or as charge frustration producers. He and Cava [14] have recently shown, that for M = Mn, Fe, and Ni, "inhomogeneous" ferromagnetic materials are formed [14-15]. In these $CaRu_{1-x}M_xO_3$ systems, $T_C$ depends also only on the dopant ion, and not on its concentration. It is thus assumed, that small ferromagnetic clusters with an intrinsic $T_C$ (depending on M) are formed, and they increase in size and volume fraction with increasing dopant concentration. Our magnetic and MS show [11], that $T_C$ of $CaRu_{1-x}Fe_xO_3$ (up to x=0.15) is similar to that of pure $CaRuO_3$, however the Fe induces a ferromagnetic structure . Charge compensation, which is assumed to be present upon $Fe^{3+}$ substitution, induces the appearance of low-spin $Ru^{5+}$ ions, which reduce the paramagnetic moment. In a recent paper [16], we have used $Cu^{2+}$($3d^9$, S=1/2) as substitution for $Ru^{4+}$, since $Cu^{2+}$ represents both charge and spin frustration. This substitution was motivated by the discovery of coexistence of magnetism and superconductivity in ruthenium-based $R_{2-x}Ce_xSr_2RuCu_2O_{10-\delta}$ and $RSr_2RuCu_2O_8$ (R=Eu, Gd) [17-18] cuprates.

We present here a magnetic study of the magnetic properties of ceramic $Ca_{1-x}M_xRuO_3$ (M=La and Sr). Indeed, several studies have appeared which probe the magnetic properties of the $Ca_{1-x}Sr_xRuO_3$ system [6,9,12,19-21]. The initial key point in all these papers is the assumption that $CaRuO_3$ and samples with 0.2<x are paramagnetic down to low temperatures. The motivation to re-investigate the $Ca_{1-x}Sr_xRuO_3$ system is based on our new understanding of $CaRuO_3$ discussed above. We are presenting here, only new results, which were not published. Using the same token, to study the magnetism further, and to test the effect of $La^{3+}$ the $Ca^{2+}$ site, which inevitably represents both charge and spin frustration disorder, results on a series of $Ca_{1-x}La_xRuO_3$ (x=0.1 and 0.5) perovskites are reported. This system is considered to be paramagnetic above 78 K and was not investigated at the low-temperatures region. Then, results of the study on $Ca_{0.8}Sr_{0.2}RuO_3$ and on $Ca_{1-x}Sr_xRuO_3$ $0.4 \leq x \leq 1$ are reported.

## 2. Experimental Details

Polycrystalline samples of $Ca_{1-x}M_xRuO_3$ (M=La and Sr), were prepared by solid-state reaction from the appropriate stoichiometric mixtures of $CaCO_3$, $SrCO_3$, $RuO_2$ (or Ru) and $La_2O_3$. Pressed pellets were preheated at 1000°C for 24 h, and then sintered at 1300-1350°C for 72h in air, with two intermediate grindings. Powder X-ray diffraction (XRD) measurements confirmed the purity of the compounds. Magnetic dc measurements were performed in a Quantum Design superconducting quantum interference device magnetometer (SQUID).

## 3. Experimental Results

$CaRuO_3$ is not paramagnetic, but rather shows the characteristics of short-range magnetic interaction, as stated above. For the sake of brevity and clarity, we present here the data accumulated on ceramic $CaRuO_3$. Fig. 1 shows the irreversibility below $T_{irr}$ =87 K, of the ZFC and FC branches measured at field of 50 Oe. As the field is increased, $T_{irr}$ is shifted to lower temperatures and the irreversibility disappears at H=10 kOe. Fig. 1 (inset) displays the almost linear isotherm magnetization at 5 K, which on an expanded scale, shows a small hysteresis loop with a coercive field of $H_C$~100 Oe and a remnant moment of $M_{rem}$=1.6 emu/mol. This remnant moment disappears at 87 K (our definition for $T_C$) [3].

### 3.1 Magnetic properties of $Ca_{1-x}La_xRuO_3$ (x=0.1 and 0.5)

The orthorhombic $LaRuO_3$ (space-group: Pbnm) can be prepared only under high oxygen pressure. However, at ambient pressure substitution up to 50% $La^{3+}$ for $Ca^{2+}$ $Ca_{1-x}La_xRuO_3$ is possible, retaining single-phase structure. $La^{3+}$ in $Ca_{1-x}La_xRuO_3$ inevitably causes both charge and lattice disorder. XRD studies confirm the orthorhombic structure (space-group: Pnma), with no secondary phases detected. The larger ionic radii of $La^{3+}$ (1.17 Å) as compared to (1.14 Å) for $Ca^{2+}$, leads to a small increase in all unit-cell parameters. The refined lattice parameters for ceramic $Ca_{1-x}La_xRuO_3$ samples are given in Table 1.

ZFC and FC magnetization curves (measured at 500 Oe) for the two $Ca_{1-x}La_xRuO_3$ (x=0.1 and 0.5) samples are shown in Fig. 2-3. The two branches merge at $T_N$ =58 (1) K and 19(1) K, respectively regardless of the applied field. No other anomalies were observed at higher temperatures. In fact, for x=0.1, the two curves do not merge completely at 58 K but rather around 90 K, indicating that a small fraction (not detectable by XRD) of $CaRuO_3$ is present. In contrast to $CaRuO_3$ (see Fig.1 inset), the isothermal magnetization curves at 5 K (up to 50 kOe) for both samples are linear, without any irreversibility (Fig.3 inset). This indicates that: (i) both samples are anti-ferromagnetically ordered and that (ii) $T_N$ decreases with increasing $La^{3+}$ concentration. It that sense, this system behaves differently from the $CaRu_{1-x}Ti_xO_3$, in which the substitution is made in the Ru-site, and $T_C$=34 K is obtained for all samples, regardless of x [11-12].

Above $T_N$, the M(T) curves of $Ca_{0.9}La_{0.1}RuO_3$ are almost linear. However, for $Ca_{0.5}La_{0.5}RuO_3$, the M(T) curves show the typical paramagnetic shape and adhere to the Curie-Weiss (CW) law: $\chi = \chi_0 + C/(T-\theta)$, where $\chi_0$ is the temperature independent part of $\chi$, C is the Curie constant, and $\theta$ is the CW temperature. A fit of the CW relation in the range of 30<T<270 K yields the values: $\chi_0$=-7 x $10^{-5}$ emu/mol Oe, $\theta$=-130(3) K and the paramagnetic effective moment $P_{eff}$ =2.7(1) $\mu_B$/Ru. Both $\theta$ and $P_{eff}$ values are similar to these obtained for pure $CaRuO_3$ (Table 1). $P_{eff}$ =2.7(1) $\mu_B$/Ru is somewhat smaller than the expected value according to Hund's rule for $Ru^{4+}$ ($4d^4$) in the low spin (S=1) state (2.83 $\mu_B$/Ru).

Since the end member $LaRuO_3$ ($Ru^{3+}$, S=0.5) is paramagnetic down to 4.2 K [7-8], we associate the decrease of $T_N$ with x in $Ca_{1-x}La_xRuO_3$ to the magnetic dilution of the $Ru^{4+}$ sublattice. The substitution of $La^{3+}$ for $Ca^{2+}$ produces charge frustration in the system and therefore requires partial reduction of neighboring $Ru^{4+}$ to lower valence state $Ru^{3+}$. The long-range three-dimensional Ru-O-Ru network in $CaRuO_3$, is disrupted by disorder. The charge compensation mechanism leads to the decrease of the net magnetic moment of the Ru site and to the systematic decrease in the magnetic transitions. However, the similarity of the paramagnetic constants (Table 1), needs some more consideration.

### 3.2 The $Ca_{1-x}Sr_xRuO_3$ system

Shown in Fig. 4 are the refined lattice parameters and the unit cell volumes for the $Ca_{1-x}Sr_xRuO_3$ system. While the a axis remains essentially unchanged, the b and c lattice parameter, and as a result, the unit cell volume, show an expansion on the increasing of Sr content. This reflects the larger ionic radius of Sr. These results compare well with data reported in Ref. 6.

### 3.2.1 *Magnetic properties of $Ca_{0.8}Sr_{0.2}RuO_3$*

Fig. 5 shows the ZFC and FC plots of $Ca_{0.8}Sr_{0.2}RuO_3$ measured at 50 Oe. Apart from the magnetic irreversibility, the main effect to be seen is the paramagnetic-like shape of both curves. Probably, this behavior has led in the past to the assumption that this material is paramagnetic [6,21]. The two curves merge at $T_{irr}$ =107 K. $T_{irr}$ is field dependent, and shifted to lower temperatures with the applied field, (Fig.5 inset). Isothermal M(H) measurements at various temperatures have been carried out. Generally speaking, all curves below $T_C$, are strongly dependent on the field (up to 7-10 kOe), until a common slope is reached. M(H) can be described as: M(H)= $M_{sat}$+ $\chi$H, where $M_{sat}$ (saturation moment) corresponds to the FM contribution of the Ru sublattice, and $\chi$H is the linear contribution to the magnetization. $M_{sat}$ which is 225 emu/mol (0.04$\mu_B$) at 5 K, decreases with temperature and becomes zero at 107 K. At low applied fields, the M(H) curve exhibits a typical ferromagnetic-like hysteresis loop (Fig. 6). Two other characteristic parameters of the hysteresis loops (at 5 K) are the remnant moment, ($M_{rem}$ = 90 emu/mol) and the coercive field ($H_C$ = 450 Oe). This $M_{rem}$, is much larger than $M_{rem}$ = 1.6 emu/mol obtained for $CaRuO_3$. $M_{rem}$ decreases gradually with the temperature and disappear at $T_C$ =107 K (Fig.6 inset).

Above $T_C$, the M(T) curves adhere closely to the CW law. A fit of in the range of 130<T<270 K yields the values: $\chi_0$ =-1.5 x $10^{-5}$ emu/mol Oe, $\theta$= –70(1) K and $P_{eff}$ =2.80(1) $\mu_B$ (Table 2). This $P_{eff}$ value fits perfectly with 2.83 $\mu_B$/Ru expected for $Ru^{4+}$ ($4d^4$) in the low spin (S=1) state. The negative $\theta$ value, is much more positive than that of $CaRuO_3$. The paramagnetic shape of the ZFC and FC curves (Fig. 5) and the negative $\theta$ obtained, suggest basically an AFM ordering for $Ca_{0.8}Sr_{0.2}RuO_3$. The hysteresis loop and the small $M_{sat}$= 0.04 $\mu_B$ as well as the irreversibility phenomena, may be a result of an anti-symmetric exchange coupling of the Dzyaloshinski-Moria (DM) type between neighboring Ru moments, which induces a local slight distortion in the Ru sublattice. The applied field causes the adjacent Ru spins to cant slightly out of their original direction and to align a component of the moment with the direction of the field. Thus, we assume that $Ca_{0.8}Sr_{0.2}RuO_3$ is a canted AFM material. In that sense its magnetic properties are quite different from the rest of the $Ca_{1-x}Sr_xRuO_3$ system discussed hereafter.

### *3.2.2 Magnetic properties of $Ca_{1-x}Sr_xRuO_3$ $0.4 \leq x \leq 1$*

Many publications have appeared already which provide an extensive study on the magnetic properties of this system [6.9.12.19-21]. The magnetic behavior of those materials is attributed to the highly correlated Ru 4d-electron band. $SrRuO_3$ (x=1) is ferromagnetically ordered and the suppression of ferromagnetism interaction for 0.4≤ x < 1, may be basically related to the orthorhombic structural, namely, the Ru-O-Ru bond angle. The difference in size between Sr and Ca (in the A site) is a dominant factor of this disorder. As stated above, this paper is limited to experimental results, which to our best knowledge, have not been reported in the past.

ZFC and FC magnetization curves for $Ca_{0.4}Sr_{0.6}RuO_3$ and $Ca_{0.2}Sr_{0.8}RuO_3$ are exhibited in Figs. 7-8. In both figures an irreversibility phenomenon is observed which it is strongly field dependent. The linear variation of $T_{ier}$ in a semi logarithmic scale for x=0.6 is shown in Fig. 7 (inset). Similar behavior has been obtained for all other series members. The ZFC shows a broad peak around 90 K, which shifts to lower temperatures for higher applied fields. The FC curve has the typical FM shape and resembles to the FM features of $Ca_{0.2}Sr_{0.8}RuO_3$ (Fig. 8) and $SrRuO_3$ (not shown). Note the difference in both the ZFC and FC curves for $Ca_{0.4}Sr_{0.6}RuO_3$ (Fig. 7) as compared to $Ca_{0.8}Sr_{0.2}RuO_3$ shown in Fig. 5. The $T_C$ and in $Ca_{1-x}Sr_xRuO_3$, increase with x as expected (Table 2).

The isothermal field dependence of the magnetization, has been measured at various temperatures and the typical FM hysteresis loops at 5 K are shown in Fig. 9. The $M_{sat}$ values listed in Table 2, are the values extrapolated to H=0 from the high field regions. The trend in $Ca_{1-x}Sr_xRuO_3$ is that: (i) $M_{sat}$ increases and (ii) $H_C$ decreases with x. The large $H_C$=9.5(1) kOe obtained for x=4, is comparable with 9 kOe observed in $CaRu_{0.75}Mn_{0.25}O_3$ [14], but much smaller than in the large $H_C$ obtained (at 5 K) in $CaRu_{1-x}Cu_xO_3$ single crystals (22 kOe) [16], and in $CaRu_{1-x}Fe_xO_3$ (<30 kOe) [11].

All samples show CW behavior in the high temperature regime and the paramagnetic parameters (Table 2) are the fit for M/H curves measured at 10 kOe, in the temperature interval from about 30 K above $T_C$ up to 325 K. Fits to the CW law were less good near $T_C$ because of short-range order and fluctuation effects. The monotonic increase of $T_C$, is followed by the increase of the θ, which are all positive as expected for ferromagnetic materials. This is consistent with previous studies on $Ca_{1-x}Sr_xRuO_3$, and has been interpreted as being due to the effect of orthorhombic distortion on the magnetic properties [9]. On the other hand the paramagnetic $P_{eff}$ values do not increase with x, but rather show a minimum for x=0.6. This unusual behavior differs from the data in Ref 6 and 19, in which $P_{eff}$ follows the general trend of $T_C$ and θ. Note, that the $M_{sat}$ values (e.g. 0.28 $\mu_B$/Ru for $Ca_{0.4}Sr_{0.6}RuO_3$ and 0.85 $\mu_B$/Ru, for $SrRuO_3$) are significantly reduced as compared to the paramagnetic moments and are much smaller than the expected ferromagnetic 2 $\mu_B$/Ru. This has been supposed to be mainly due to itinerancy of the magnetic 4d electrons [6,9].

The main effect to be seen is the temperature dependence of $H_C$ shown in Figs. 10-11. For x=0.4 and 0.6, $H_C$ decreases rapidly with T and then increase with a peak at 90 and 115 K respectively (Fig. 10 inset) and becomes zero at $T_C$. On the other hand, $H_C$ for $SrRuO_3$, decreases slightly up to 50 K and exhibits a peak around 90 K. Since the common structural feature of all compounds is that they are composed of an array of corner-shared $RuO_6$ octahedra, it is assumed that the degree of tilting and rotation of these octahedra around 80-90 K, governs this peculiar phenomenon. The peaks in $H_C$ are unique and needs more experimental studies, such as neutron diffraction measurements at low temperatures.

**4. Summary**

$CaRuO_3$, is not paramagnetic, but rather shows the characteristics of short-range magnetic interactions, possibly spin-glass behavior with $T_C$= 87 K. The $La^{3+}$ substitution for $Ca^{2+}$ in $Ca_{1-x}La_xRuO_3$, induces a disorder in the A site and produces charge frustration. This disorder, has a significant yet unexpected effect on the magnetic properties of $CaRuO_3$. The system becomes antiferromagnetically ordered and the magnetic transitions shift to 58 and 19 K for x=0.1 x=0.5 respectively.

The main effects to be seen in the $Ca_{1-x}Sr_xRuO_3$ system are: (i) $T_C$ of $CaRuO_3$ is shifted monotonically to higher temperatures with Sr. (ii) $Ca_{0.8}Sr_{0.2}RuO_3$ orders at 107 K as a canted antiferromagnetically system (iii) The compounds with higher Sr concentration display a ferromagnetic behavior and the saturation moment increases with Sr. Yet, the value for $SrRuO_3$ is smaller that the expected one for $Ru^{4+}$ (S=1). (iv) The coercive field is larger for the x=0.4 compound (9.5 kOe) and decreases with Sr, (2.4 kOe for $SrRuO_3$). (v) $H_C$ for x=0.4 and 0.6 decreases sharply with temperature and than increases again with a peak at 95 and 110 K, respectively. (vi) $H_C$ for $SrRuO_3$, does not change much at low temperatures, shows a peak at 90 K and then decrease sharply and becomes zero at $T_C$ =165 K.

**Acknowledgment**: We gratefully acknowledge the support from the BSF (1999).

**References.**

[1] T. C. Gibb, R. G. Greatrex, N.N. Greenwood and P. Kaspi, J. Chem. Soc. Dalton Trans. (1973) 1253.

[2] L. Klein, L. Antognazza, T.H. Geballe, M.R. Beasley, A. Kapitulnik, Phys. Rev. B 60 (1999) 1448

[3] I. Felner, I. Nowik, I. M. Bradaric and M. Gospodinov, Phys. Rev.B 60 (2000)1448.

[4] J. Lynn (unpublished results)

[5] P. A. Cox , R.G. Egdell, J.B. Goodenough, A. A. Hamnett and C. C. Naish, J. Phys. C: Solid State Phys. 16 (1983)6221.

[6] G. Cao, S. McCall, M. Shepard, J.E. Crow and R.P. Guertin, Phys. Rev. B 56 (1997) 321.

[7] R. J. Bouchard and J.F. Weiher, J. Solid State Chem. 4, (1972) 80.

[8] T. Sugiyama and N. Tsuda, J. Phys. Soc. Japan 68 (1999) 3980.

[9] M. Shepard, G. Cao, S. McCall, F. Freibet and J.E. Crow, J. Appl. Phys 79 (1996) 4821.

[10] G. Cao, F. Freibert and J.E. Crow, J. appl. Phys. 81 (1997) 3884.

[11] I. Felner, U. Asaf, I. Nowik and I. Bradaric, Phys. Rev. B (2002) in press

[12] T. He and R.J. Cava, Phys. Rev. B 63 (2001) 172403.

[13] G. Cao, S. McCall, J. Bolivar, M. Shepard, F. Freibert , P. Henning and J.E. Crow, Phys. Rev. B 54 (1996)15144.

[14] T. He and R.J. Cava, J. Phys.: Condens. Matter 13 (2001) 8347.

[15] A. Maignan, C. Martin, M. Hervieu and B. Raveau, Solid State Commun.117 (2001) 377.

[16] I.M. Bradaric, M. Gospodinov and I. Felner, Phys. Rev.B, 65 (2002) 024421.

[17] I. Felner, U Asaf, Y. Levi and O. Millo, Phys. Rev. B 55 (1997) R3374.

[18 J.L.Pringle, J.L.Tallon, B.G Walker and H.J. Trodahl, Phys. Rev. B 59 (1999) R11679.

[19] T. He, Q. Huang  and R.J. Cava, Phys. Rev. B 63 (2000) 024402.

[20] K. Yoshimura, T. Imai, T. Kiyama, K.R. Thurber, A.W. Hunt and K. Kosuge, Phys. Rev. Lett. 83 (1999) 4397.

[21] T. Kiyama, K. Yoshimura, K. Kosuge, H. Michor and G. Hilscher, J. Phys. Sec. Japan 67 (1998) 307.

**Table 1**: Unit cell and magnetic parameters of $Ca_{1-x}La_xRuO_3$

| Ca$_{1-x}$La$_x$RuO$_3$ | a (Å) | b (Å) | c (Å) | V (Å)$^3$ | T$_N$ (K) | P$_{eff}$ (μ$_B$/Ru) | θ (K) |
|---|---|---|---|---|---|---|---|
| CaRuO$_3$ | 5.523 | 7.649 | 5.359 | 226.4 | 87(1) | 2,7 | -138 (5) |
| x=0.1 | 5.536 | 7.666 | 5.383 | 228.4 | 58 (1) | - | - |
| x=0.5 | 5.585 | 7.762 | 5.399 | 234.0 | 19(1) | 2,7 | -130(3) |

**Table** 2   Magnetic parameters of Ca$_{1-x}$Sr$_x$RuO$_3$.

| Ca$_{1-x}$Sr$_x$RuO$_3$ | T$_C$ (K) | M$_{sat}$ (5 K) emu/mol | H$_C$ (5 K) kOe. | χ$_0$ emu/mol Oe | P$_{eff}$ (μ$_B$/Ru) | θ (K) |
|---|---|---|---|---|---|---|
| CaRuO$_3$ | 87 (1) | - | 0.10 | 5 x 10$^{-4}$ | 2.66 | -138 |
| Ca$_{0.8}$Sr$_{0.2}$RuO$_3$ | 107 | 225 | 0.45 | -1.5 x 10$^{-5}$ | 2.80 | -70 |
| Ca$_{0.6}$Sr$_{0.4}$RuO$_3$ | 147 | 1600 | 9.5 | 2.4 x 10$^{-4}$ | 2.51 | 47 |
| Ca$_{0.4}$Sr$_{0.6}$RuO$_3$ | 158 | 2835 | 9.0 | 4 x 10$^{-4}$ | 2.38 | 106 |
| Ca$_{0.2}$Sr$_{0.8}$RuO$_3$ | 163 | 4200 | 3.4 | -7 x 10$^{-6}$ | 2.61 | 141 |
| SrRuO$_3$ | 165 | 4790 | 2.4 | -9x 10$^{-4}$ | 2.77 | 161 |

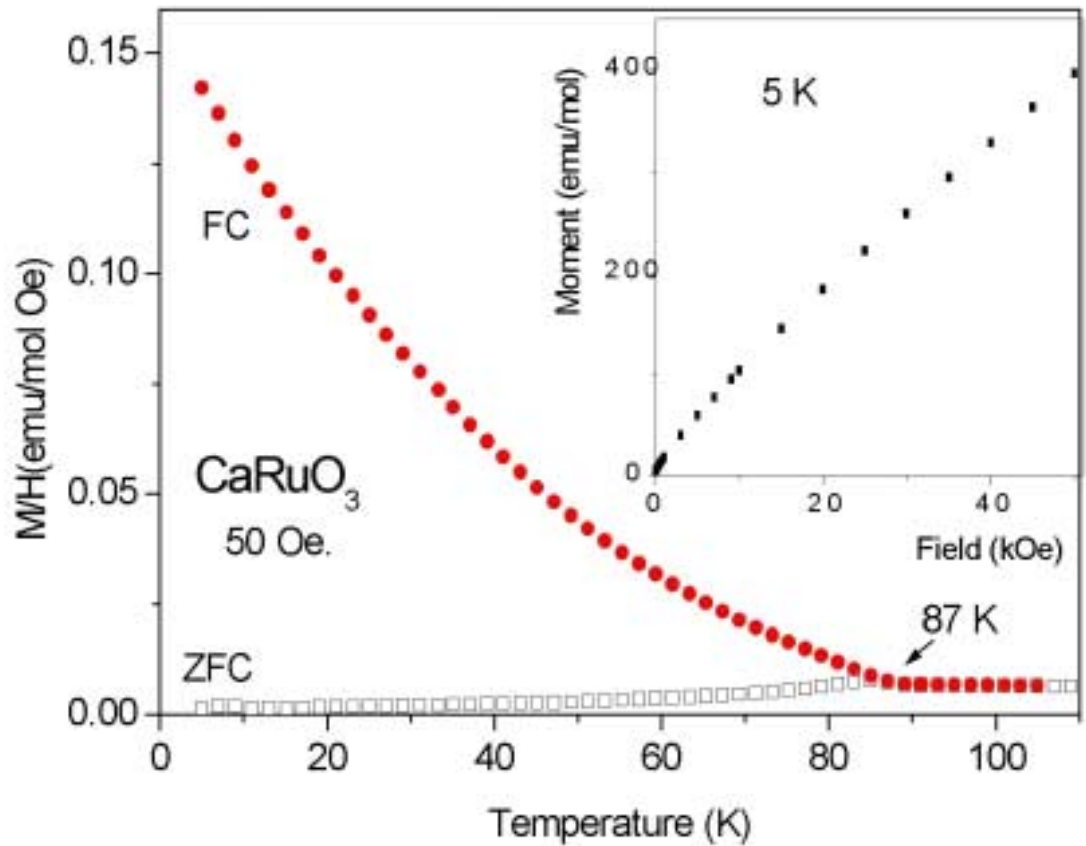

Fig. 1 ZFC and FC susceptibility studies at 50 Oe of $CaRuO_3$. The isothermal magnetization at 5 K up to 50 kOe is shown in the inset.

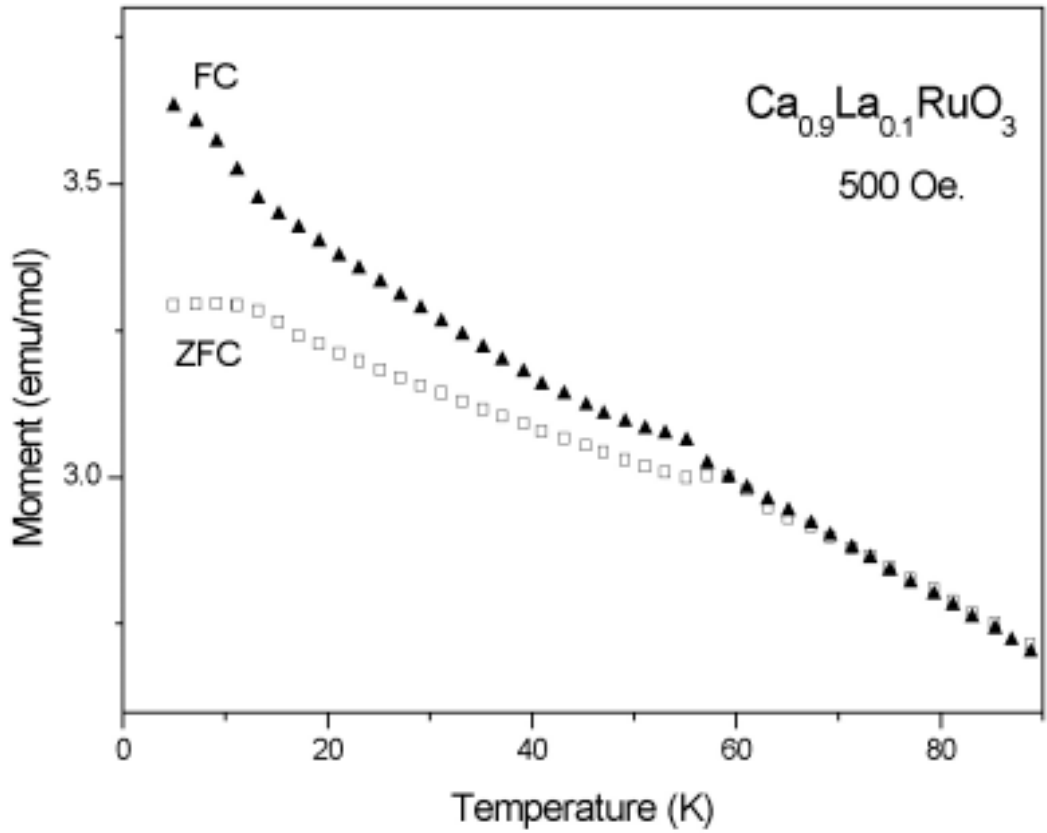

Fig. 2  ZFC and FC magnetic curves measured at 500 Oe of  $Ca_{0.9}La_{0.1}RuO_3$.

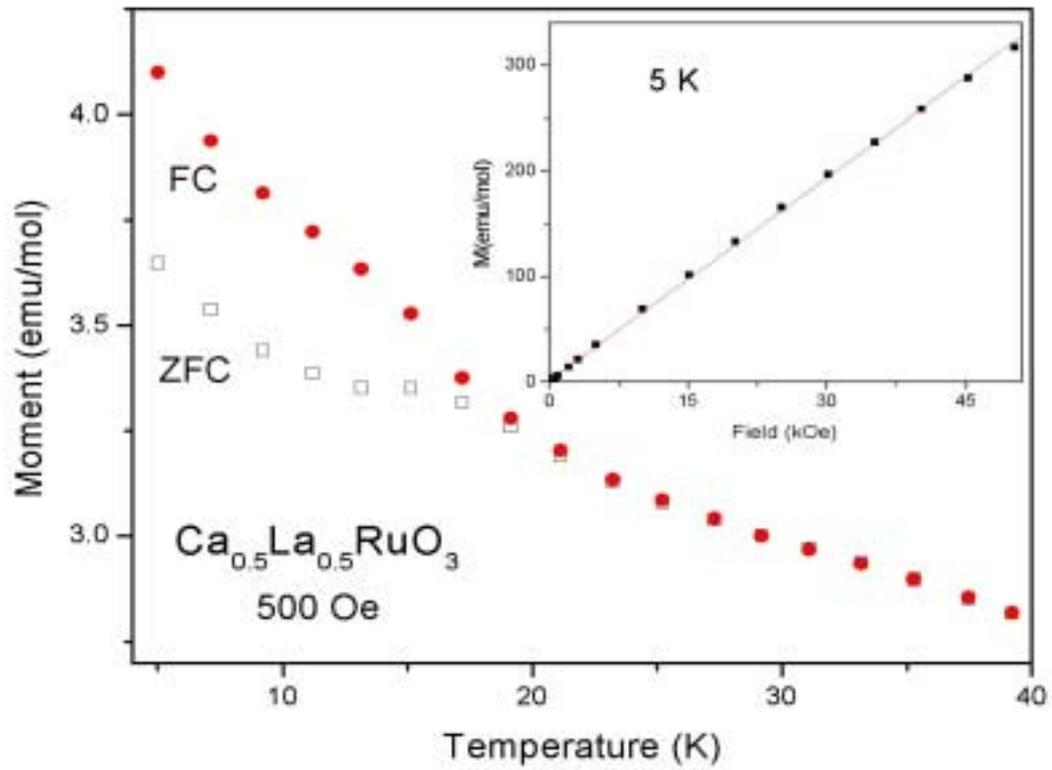

Fig. 3  ZFC and FC curves of $Ca_{0.5}La_{0.5}RuO_3$ measured at 500 Oe. The inset shows the linear isothermal magnetization at 5 K

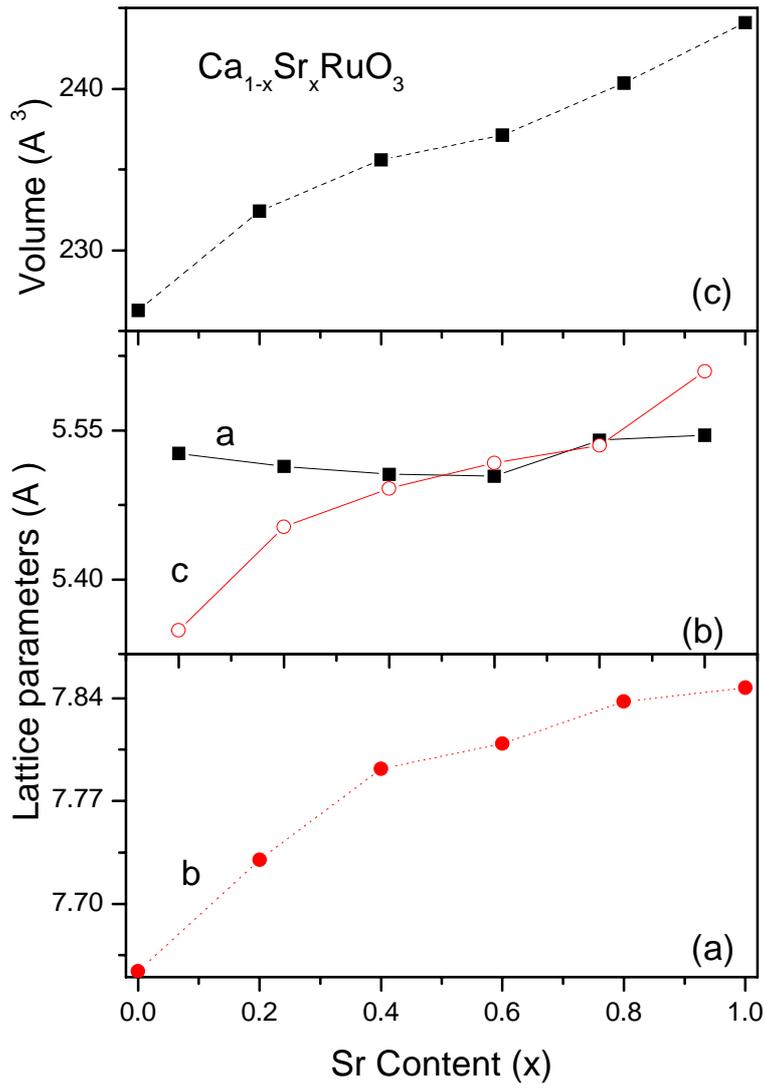

Fig. 4 The lattice parameters and the unit cell volume of $Ca_{1-x}Sr_xRuO_3$.

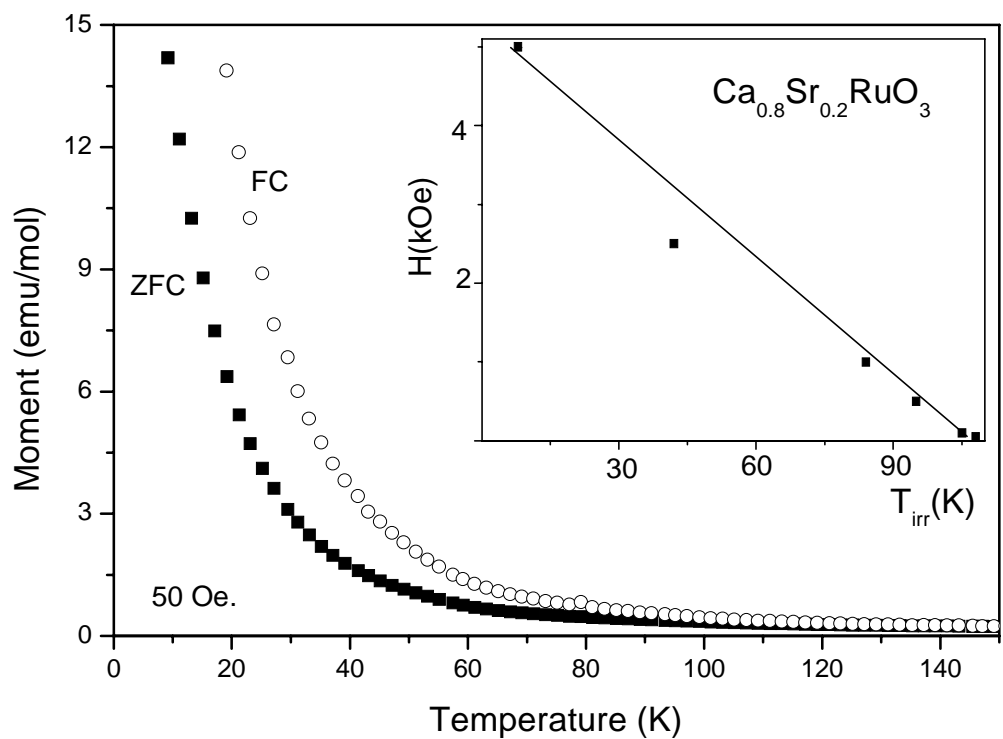

Fig. 5. ZFC and FC magnetic curves of $Ca_{0.2}Sr_{0.8}RuO_3$ measured at 50 Oe. The inset shows the temperature dependence of the irreversibility temperature.

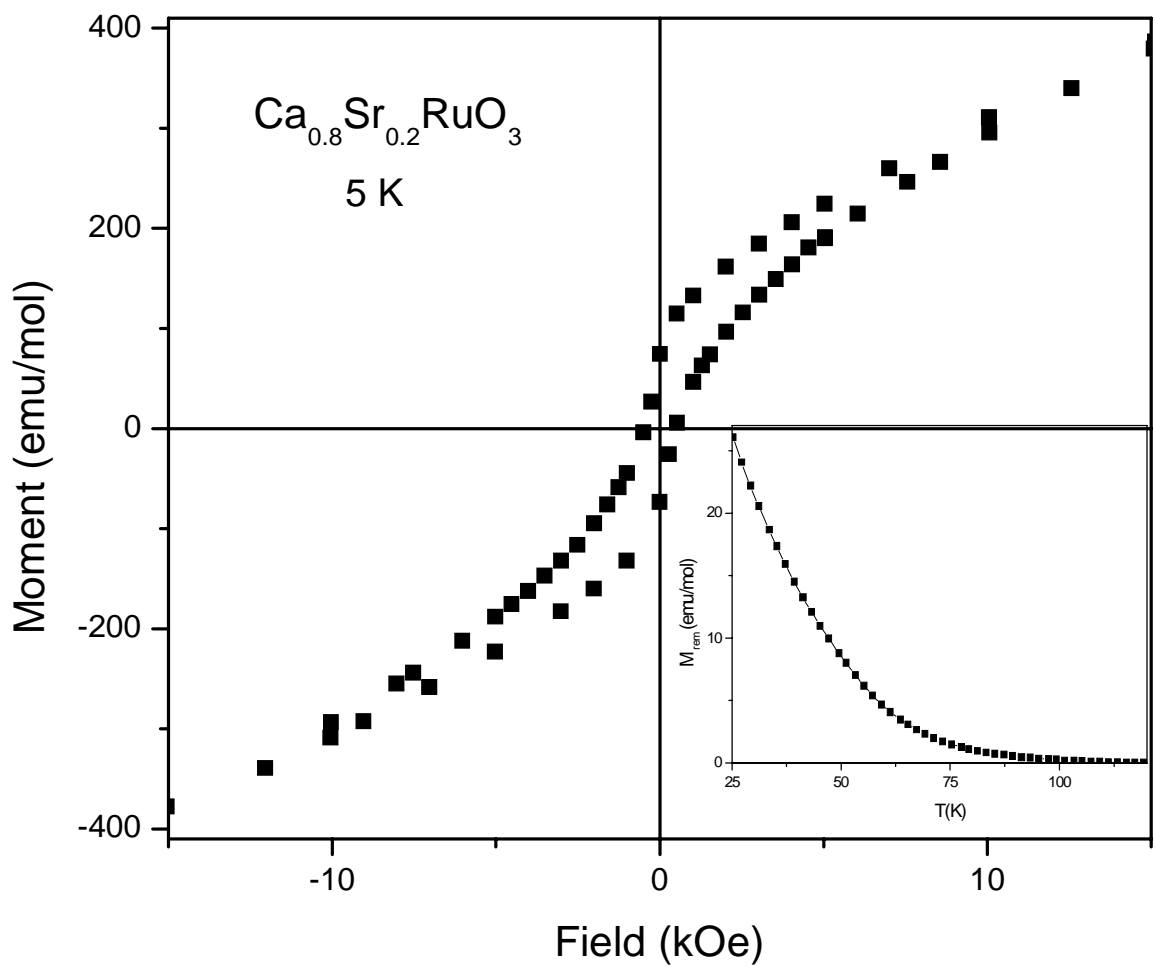

Fig. 6 Hysteresis loops at 5 K up to 15 kOe of $Ca_{0.2}Sr_{0.8}RuO_3$. The inset shows the variation of the remnant magnetization (obtained at 5) with temperature. .

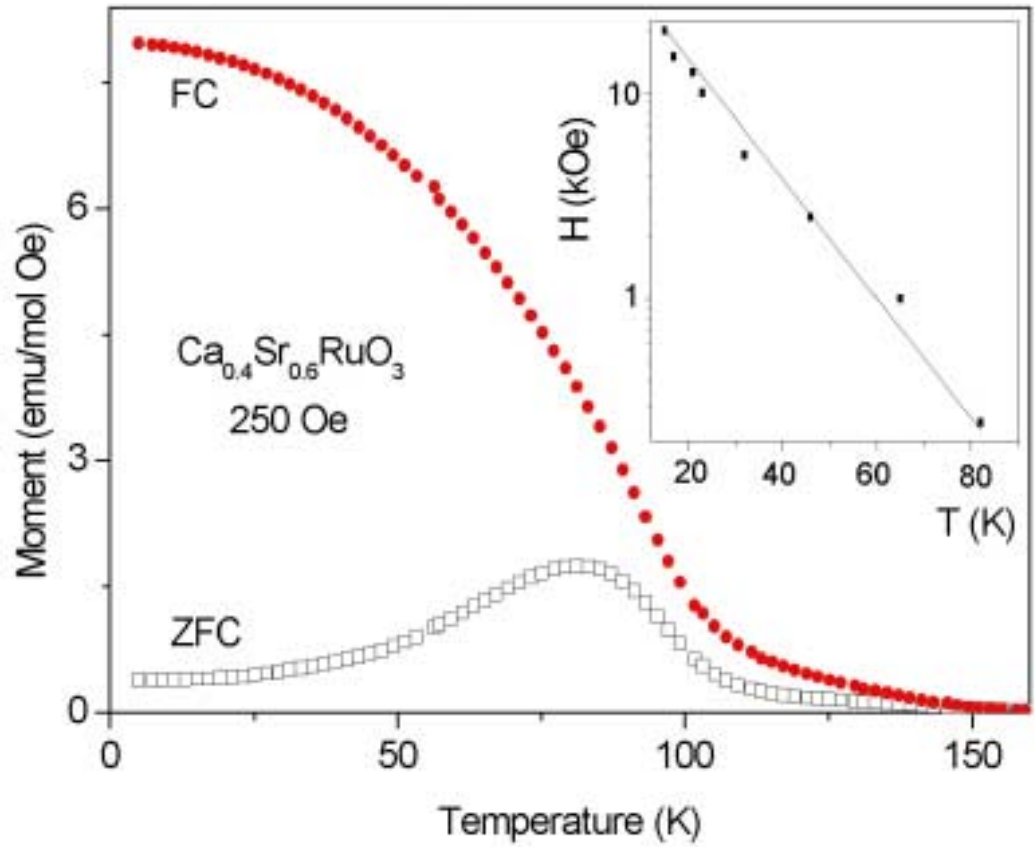

Fig. 7. ZFC and FC magnetic curves of $Ca_{0.4}Sr_{0.6}RuO_3$ measured at 250 Oe. The inset shows the temperature dependence of the irreversibility temperature. The solid line is a fit to a linear relation between $T_{irr}$ and $\ln H$.

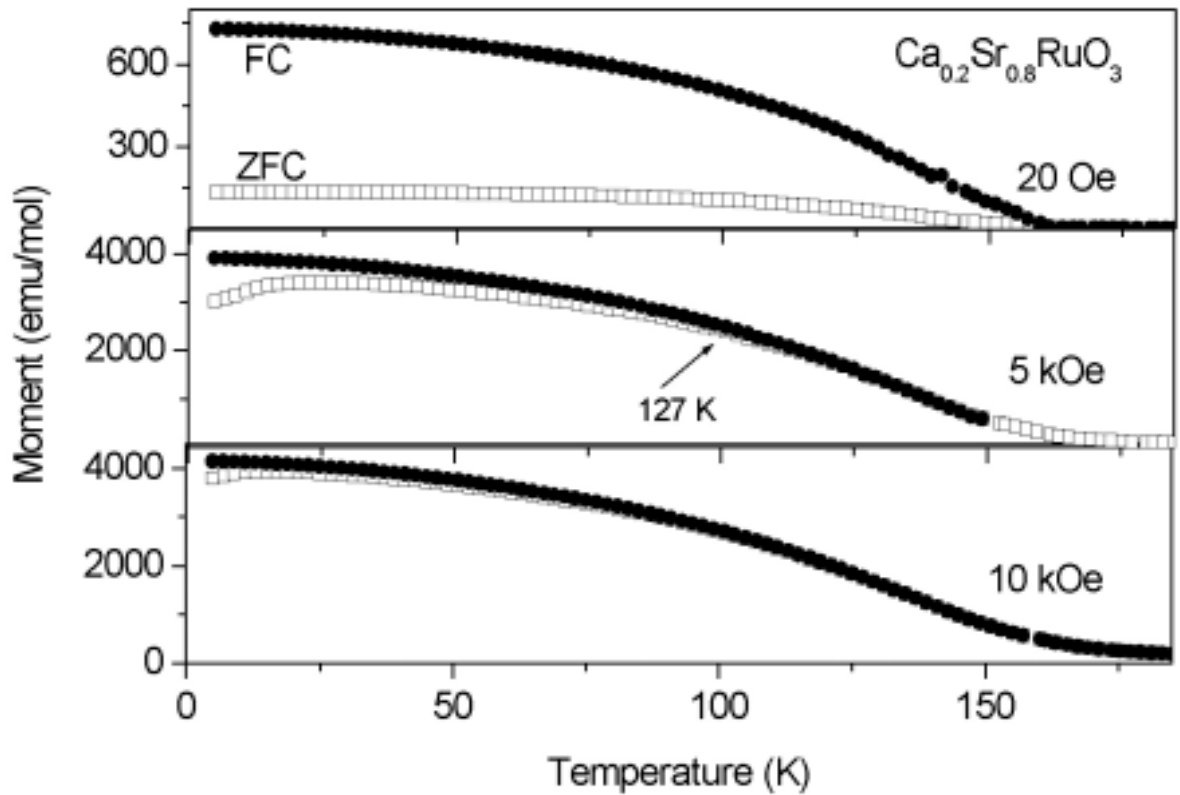

Fig. 8. ZFC and FC magnetic curves of $Ca_{0.2}Sr_{0.8}RuO_3$ measured various applied fields. For H=10 kOe $T_{irr}$ =104 K.

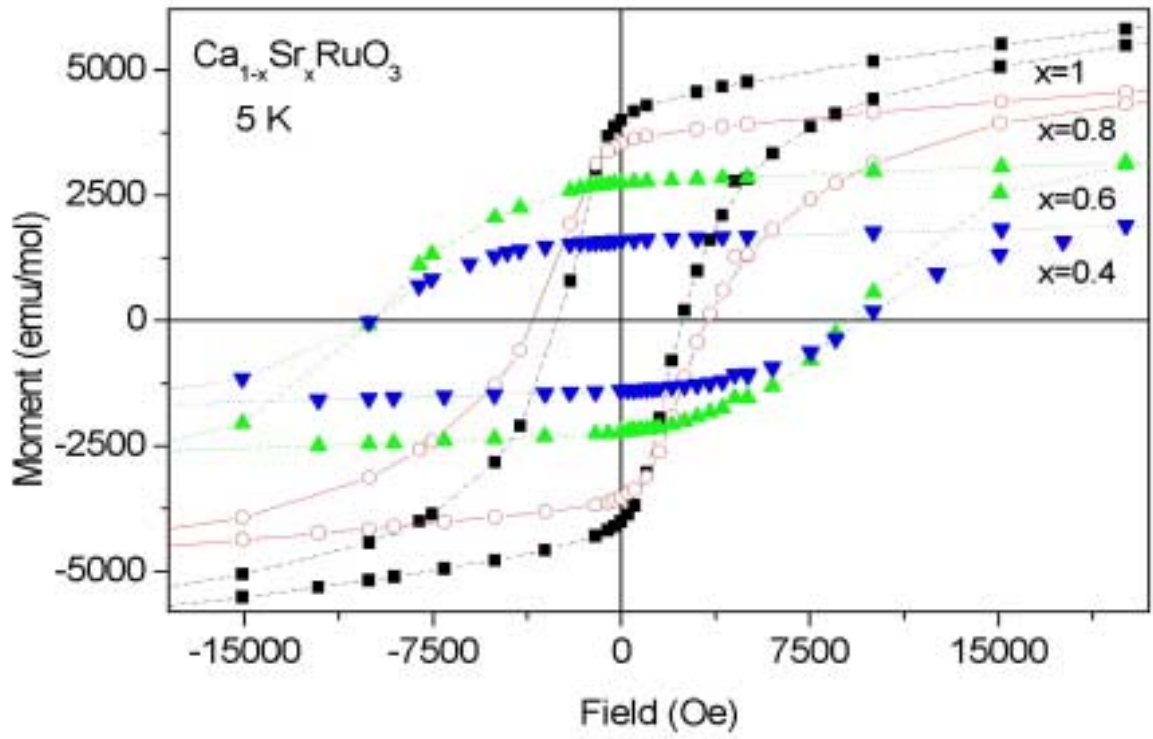

.Fig. 9. Hysteresis loops for $Ca_{1-x}Sr_xRuO_3$

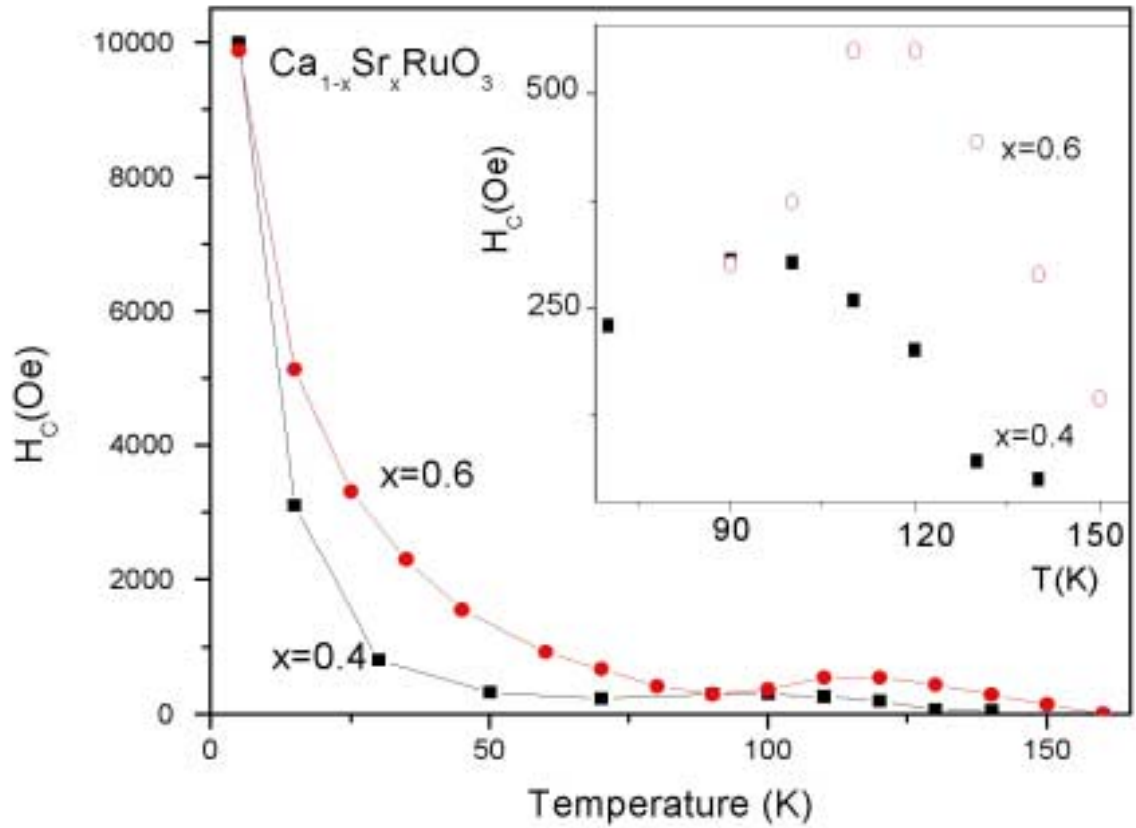

.Fig 10.. The temperature dependence of the coercive fields for $Ca_{1-x}Sr_xRuO_3$ (x=0.4 and 0.6). Note the extended scale of the inset

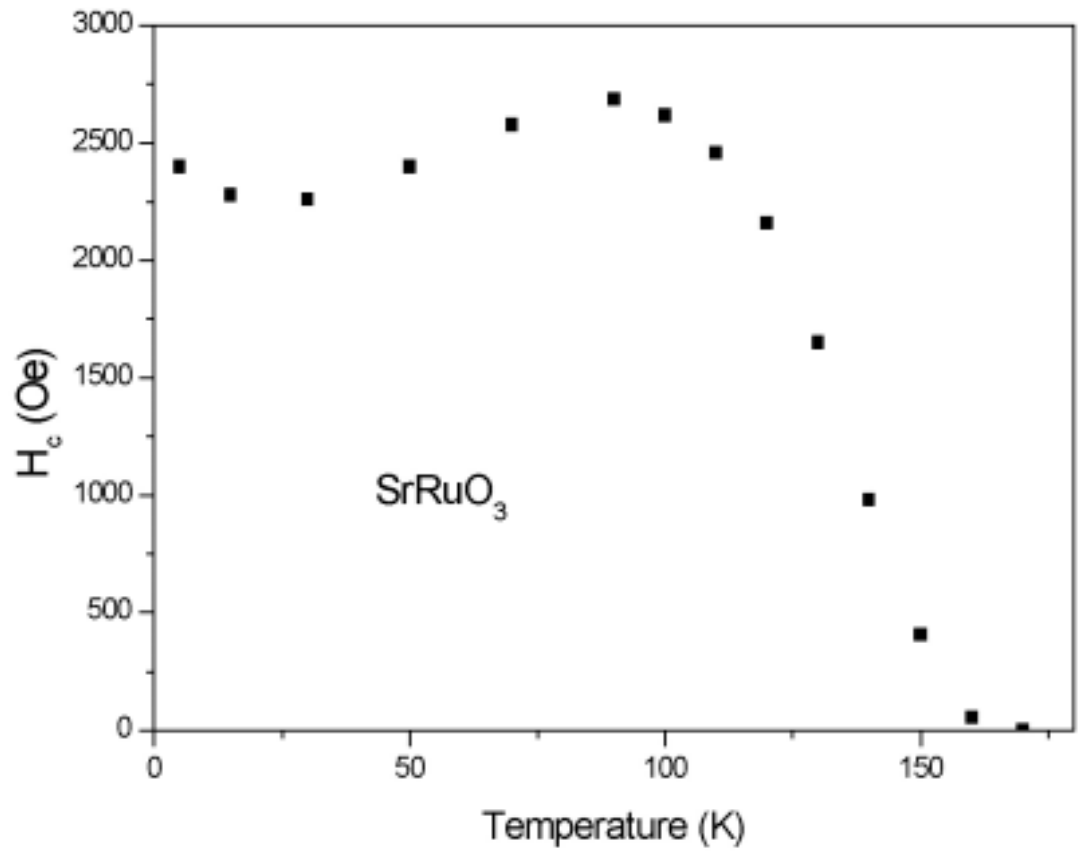

Fig 11. The temperature dependence of the coercive fields for $SrRuO_3$
.